\documentclass{aims}

\usepackage{amsmath}
  \usepackage{paralist}
  \usepackage{graphics} 
  \usepackage{epsfig} 
 \usepackage[colorlinks=true]{hyperref}

\hypersetup{urlcolor=blue, citecolor=red}

\def\currentvolume{X}
 \def\currentissue{X}
  \def\currentyear{200X}
   \def\currentmonth{XX}
    \def\ppages{X--XX}

\theoremstyle{definition}

\title[Vortices in the Zabusky-McWilliams Model]
      {Interactions of point vortices in the Zabusky-McWilliams model with a background flow}

\author[Colm Connaughton and John R. Ockendon]{}

\subjclass{Primary: 58F15, 58F17; Secondary: 53C35.}
 \keywords{Fluid dynamics, point vortex models}

 \email{connaughtonc@gmail.com}
 \email{ock@maths.ox.ac.uk}

\newcommand{\x}{{\mathbf x}}
\newcommand{\y}{{\mathbf y}}
\renewcommand{\v}{{\mathbf v}}
\newcommand{\pd}[2]{\frac{\partial #1}{\partial #2}}
\newcommand{\dd}[2]{\frac{d #1}{d #2}}

\begin{document}
\maketitle

\centerline{\scshape Colm Connaughton }
\medskip
{\footnotesize
 \centerline{Warwick Mathematics Institute and Warwick Centre for Complexity Science}
   \centerline{University of Warwick}
   \centerline{Coventry, CV4 7AL, United Kingdom}
} 

\medskip

\centerline{\scshape John R. Ockendon}
\medskip
{\footnotesize
 \centerline{Oxford Centre for Industrial and Applied Mathematics }
   \centerline{Mathematical Institute, University of Oxford,}
   \centerline{24--29 St Giles', Oxford OX1 3LB, United Kingdom}
}

\bigskip

\begin{abstract}
We combine a simple quasi-geostrophic flow model with the Zabusky-McWilliams theory of
atmospheric vortex dynamics to address a hurricane-tracking problem of interest to the
insurance industry. This enables us to make predictions about the ``follow-my-leader''
phenomenon.
\end{abstract}

\section{Introduction} 
\label{sec-intro}

\subsection{Insurance industry motivation for the ``follow-my-leader'' problem}
The 2005 Atlantic hurricane season is famous as the most active and expensive Atlantic
hurricane season since records began.  In addition to thousands of deaths, damage to property
and infrastructure was estimated to have amounted to 130 billion USD.  The Mexican states
of Quintana Roo and Yucat\'{a}n and the U.S. states of Florida and Louisiana were each struck
twice by large hurricanes.  These included Katrina, the most expensive natural disaster in the
history of the United States. The insurance industry is naturally interested in the question
of whether this is mere coincidence or whether there are correlations between the
tracks of intense hurricanes. Risk estimation models used in the industry often treat the
probability of hurricanes making landfall in a particular area as independent Poisson
processes characterised by their historical mean. The possibility of even weak correlations,
particularly between large storms, may be an important source of systematic error in
these models.

This question was posed by Lloyds and explored during the 73rd European Study Group with
Industry which took place at Warwick in April 2010 \cite{ESGI73}. Statistical analysis of the historical
data was done which did not prove conclusive. In tandem with this statistical analysis a
more basic fluid dynamics question was studied: how do the pairs of vortices embedded in a
larger scale ``steering flow'' influence each other when their separations are large? 

Our work will build on the basic theories of oceanic-scale atmospheric flows described
in \cite{CHA1949,PED1987} and the theories of hurricane dynamics proposed in \cite{BTK2007,BEYL1998,VFVM2003}. We will rely heavily on the atmospheric vortex dynamics model proposed
in \cite{ZMcW1982} and the observational studies reported in \cite{LH1993}.

\subsection{Inviscid vortex dynamics in 2-D: summary of the results of the Study Group}

The most elementary theory of 2-D vortex dynamics concerns the flow that results when vortices of strengths
$\Gamma_i$ at $\x_i(t) = (x_i(t), y_i(t))$, $i=1\ldots n$, move in a background potential flow,
\begin{displaymath} 
\mathbf{U}_0 = \left(\pd{\phi_0}{x}, \pd{\phi_0}{y}\right) = \left(-\pd{\psi_0}{y}, \pd{\psi_0}{x}\right),
\end{displaymath}
where $\psi_0(x,y,t)$ is a prescribed streamfunction. The theory asserts that the streamfunction is
\begin{equation}
\label{eq-streamfunctionEuler}
\psi(x,y,t) = \psi_0(x,y,t) + \sum_{i=1}^{n} \frac{\Gamma_i}{2\,\pi}\,\log r_i,
\end{equation}
where $r_i^2 = (x-x_i)^2 + (y-y_i)^2$, and that the $(x_i(t), y_i(t))$ evolve according to the Hamiltonian
system:
\begin{eqnarray}
\nonumber \dd{x_i}{t} &=& -\left.\pd{\psi_0(\x)}{y}\right|_{\x=\x_i(t)} - \sum_{j=1}^n \frac{\Gamma_j(t)\,\left( y_i(t)-y_j(t)\right)}{2\,\pi\,r_{ij}^2},\\
\label{eq-2DEuler}
\dd{y_i}{t} &=& \left.\pd{\psi_0(\x)}{x}\right|_{\x=\x_i(t)} + \sum_{j=1}^n \frac{\Gamma_j(t)\,\left( x_i(t)-x_j(t)\right)}{2\,\pi\,r_{ij}^2}.
\end{eqnarray}
Eq.~(\ref{eq-streamfunctionEuler}) ensures that the flow is potential flow away from the vortices and 
Eq.~(\ref{eq-2DEuler}) is the Helmholtz condition that each vortex moves with the velocity that would have
existed in its absence. This condition can be justified by smearing the vorticity into small patches 
(see \cite{VFVM2003} for the use of this idea to model hurricanes) around $(x_i, y_i)$ and 
considering the momentum balance for these patches, which reveals that no relative velocity can exist
between the patch and its ambient free stream.

\begin{figure}
\centering
\begin{tabular}{cc}
  (A) Unperturbed track & (B) Track perturbed by a second vortex\\
  \includegraphics[width=5cm]{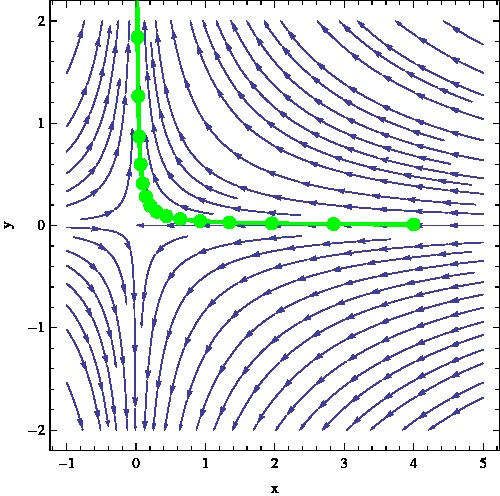}& \includegraphics[width=5cm]{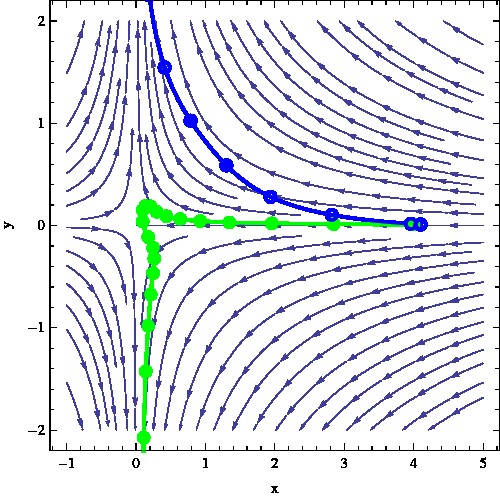}
\end{tabular}
\caption{\label{fig-deflectionEuler} Motion of point vortices in the steering flow, Eq.~(\ref{eq-potFlow}). (A) shows a single vortex following the streamlines of the steering flow. (B) 
shows the large change in trajectory  which can occur if the original vortex (solid circles)
is near a stagnation point when a second vortex (open circles) is introduced at a later time.}
\end{figure}

A very crude model for the ``follow-my-leader'' problem is to consider the motion of two equal vortices placed
in a steering flow comprising of a uniform potential flow impinging upon a solid wall 
(see Fig. \ref{fig-deflectionEuler}(A)). The flow in the upper right quadrant may be thought 
of as representing the Atlantic anticyclone and the wall as representing a blocking pattern
in the Gulf of Mexico. We took
\begin{equation}
\label{eq-potFlow}
\psi_0 = U\,x\,y,
\end{equation}
The streamlines of the large
scale steering flow are plotted in Fig.~\ref{fig-deflectionEuler}. The only interesting
feature is that the flow has a hyperbolic point at $(0,0)$. With
$\Gamma=0$ the point vortices simply follow the streamlines shown in
Fig.~\ref{fig-deflectionEuler}(A).

We performed a number of numerical experiments to show how $\Gamma \neq 0$
affects this passive advection of vortices. Fig.~\ref{fig-deflectionEuler}(B) shows the
tracks of two equal vortices having $\Gamma=1$ which start from the same point as
in Fig. \ref{fig-deflectionEuler}(A), (this was $(4.00, 0.01)$ in the figure) but separated
by a delay in time. The lead vortex follows the same path until the second vortex is
introduced. The lead vortex then undergoes a deflection which, because it occurs when the
lead vortex is near to the stagnation point, subsequently results in a completely different
southwards trajectory to the northwards trajectory it would have followed in the absence of
the second vortex. Note that the second vortex also experiences a deflection but still
follows roughly the same path.

Of course this large deviation in the trajectory of the first vortex does not occur for all 
configurations of the two vortices. The conclusion to be drawn from this simple model is, 
as one might expect, that the question of whether vortices follow each other is
not straightforward even in this simplest case. Trajectories which
come close to stagnation points of the steering flow are very difficult
to predict.

\subsection{Rotational background flow}

A serious complication arises when we try to generalise the above approach to
{\em rotational} steering flows,  $\mathbf{U}_0 = (U_0, V_0)$ where $\nabla\cdot \mathbf{U_0} =0$
and $(\mathbf{U}\cdot\nabla)\,\mathbf{U} = -\nabla p$ in dimensionless variables. Now
$\mathbf{U}_0$ is derived from a streamfunction that satisfies
\begin{equation}
J\left[\psi_0,\Delta\,\psi_0\right] = 0,
\end{equation}
where  $J\left[f,g\right]$ is the Jacobian operator,
\begin{displaymath}
J\left[f,g\right] = \pd{f}{x}\,\pd{g}{y} - \pd{f}{y}\,\pd{g}{x}.
\end{displaymath}
Considering just one moving vortex for simplicity, it is now easy to see that if we suppose that
\begin{equation}
\label{eq-streamfunctionEuler2}
\psi(x,y,t) = \psi_0(x,y,t) + \frac{\Gamma}{4\,\pi}\,\log \left[  (x-x_0)^2 + (y-y_0)^2 \right],
\end{equation}
then
\begin{displaymath}
\Delta\,\psi = \Delta\,\psi_0 + \Gamma\,\delta(x-x_0)\,\delta(y-y_0).
\end{displaymath}
Also
\begin{eqnarray*}
\pd{\psi}{y} &=& \pd{\psi_0}{y} + \frac{\Gamma}{2\,\pi}\frac{y-y_0}{(x-x_0)^2 + (y-y_0)^2}\\
\pd{\psi}{x} &=& \pd{\psi_0}{y} + \frac{\Gamma}{2\,\pi}\frac{x-x_0}{(x-x_0)^2 + (y-y_0)^2}
\end{eqnarray*}
and
\begin{eqnarray*}
\pd{\Delta\,\psi}{x} &=& \Delta\,\pd{\psi_0}{x} + \Gamma \delta^\prime(x-x_0)\,\delta(y-y_0)\\
\pd{\Delta\,\psi}{y} &=& \Delta\,\pd{\psi_0}{y} + \Gamma \delta(x-x_0)\,\delta^\prime(y-y_0)\\
\pd{\Delta\,\psi}{t} &=& \Gamma \dot{x}_0\, \delta^\prime(x-x_0)\,\delta(y-y_0) + \Gamma \dot{y}_0\, \delta(x-x_0)\,\delta^\prime(y-y_0).
\end{eqnarray*}
Hence when we collect terms in the vorticity equation,
\begin{equation}
\label{eq-vorticityEqn}
\pd{\Delta\,\psi}{t} + J\left[\psi_0,\Delta\,\psi_0\right] = 0,
\end{equation}
we find that the coefficients of $ \delta^\prime(x-x_0)\,\delta(y-y_0)$ and $\delta(x-x_0),\delta^\prime(y-y_0)$
are $\dot{x}_0 + \pd{\psi_0}{y}$ and $\dot{y}_0 - \pd{\psi_0}{x}$ respectively, in accordance with the
Helmholtz condition, Eq.~(\ref{eq-2DEuler}). However there are also terms like
\begin{displaymath}
\frac{(x-x_0)\,\delta(x-x_0)\,\delta^\prime(y-y_0)}{(x-x_0)^2 + (y-y_0)^2},
\end{displaymath}
which vanish and then terms like
\begin{displaymath}
\left[\frac{y-y_0}{(x-x_0)^2 + (y-y_0)^2}\pd{}{x} - \frac{x-x_0}{(x-x_0)^2 + (y-y_0)^2}\pd{}{y} \right]\, \Delta\,\psi_0,
\end{displaymath}
which only vanish in a potential flow. What has happened is that the introduction of the vortex has 
interfered with the global distribution of vorticity meaning that we have to solve the full vortical 
Euler equation, Eq.~(\ref{eq-vorticityEqn}) everywhere. Since almost no steering flow for hurricane
tracks can be approximated by a potential flow, this means that the results of point vortex models such as
 that presented above have very limited usefulness unless the generation of background vorticity caused by
the introduction of the point vortex is small in some sense. This may be the case if the strengths of the
moving vortices are weak so that $\frac{\Gamma}{4\,\pi}$ in Eq.~(\ref{eq-streamfunctionEuler2}) can be
replaced by a small parameter, $\epsilon$. If we try the ansatz,
\begin{equation}
\label{eq-ansatz}
\psi(x,y,t) = \psi_0(x,y,t) + \epsilon \left[ \log \left[  (x-x_0)^2 + (y-y_0)^2 \right] + \psi_1\right] + O(\epsilon^2),
\end{equation}
where $\psi_1$ is non-singular at $(x_0,y_0)$, then we find that we can make the terms of $O(\epsilon)$ in
Eq.~(\ref{eq-vorticityEqn}) if $\psi_1$ satisfies:
\begin{eqnarray*}
\pd{\psi_1}{t} + \left(\pd{\psi_0}{y}\,\Delta\,\pd{}{x} - \pd{\psi_0}{x}\,\Delta\,\pd{}{y}\right)\psi_1
+ \left(\pd{\psi_1}{y}\pd{}{x} - \pd{\psi_1}{x}\pd{}{y}\right) \,\Delta\,\psi_0 \\
= \left[\frac{y-y_0}{(x-x_0)^2 + (y-y_0)^2}\pd{}{x} - \frac{x-x_0}{(x-x_0)^2 + (y-y_0)^2}\pd{}{y} \right]\, \Delta\,\psi_0,
\end{eqnarray*}
so that the size of the perturbation can, in principle, be obtained by solving a linear equation.

The feedback from the vortex dynamics to the steering flow is even more dramatic when a
vortex of strength $\Gamma$ is introduced at the origin into a weak steering flow with
streamfunction $\epsilon \psi_0(x,y)$. Then a simple perturbation argument shows that the
streamfunction, $\epsilon \psi(,x,y,t)$ evolves according to
\begin{displaymath}
\Delta\, \pd{\psi}{t} = \frac{\Gamma}{2\,\pi (x^2+y^2)}\,\left(y\pd{}{x} - x\pd{}{y} \right)\,\Delta \psi
\end{displaymath}
on a timescale which is short compared to that of the vortex motion. Hence the vorticity
in the steering flow has to be rearranged before the vortex trajectory can be computed.

This discussion illustrates one of the main difficulties of building a more realistic model of the
steering of hurricanes by large scale atmospheric motions since such a model must necessarily incorporate the
rotation of the Earth and the background vorticity which it induces. The problems associated with the 
incorporation of the Earth's rotation can be partially addressed using a model originally devised by
Zabusky and McWilliams \cite{ZMcW1982} although we note that we are still required to assume that the 
production of background vorticity remains small. We discuss this model in the next section.

\section{Zabusky-McWilliams model of point vortices on the $\beta$-plane}

\subsection{The Charney equation}

The atmosphere is, to leading order, in geostrophic balance. That is to say, the largest
terms in the equations of motion which approximately balance each other, are the horizontal 
pressure gradient and the coriolis force induced by the Earth's rotation (see \cite{PED1987} 
for a full discussion). Any reduced model of atmospheric dynamics must take rotation into
account if it is to have any chance of being applicable. The most basic pde model of
the quasi-two-dimensional dynamics characteristic of large scale atmospheric motion are the
so-called quasi-geostrophic equations (again discussed in great detail in \cite{PED1987}),
which, in the simplest case of pure barotropic motion can be reduced to the Charney
equation \cite{CHA1949} for a single scalar streamfunction, $\Psi(\x,t)$, of a two-dimensional 
spatial coordinate, $\x = (x,y)$ and time, $t$. Written on the so-called $\beta$-plane,
where $x$ denotes the longitudinal direction and $y$ denotes the latitudinal direction,
the Charney equation is written
\begin{equation}
\label{eq-CHM}
\pd{}{t}\left(\Delta \Psi - \gamma^2\,\Psi\right) +\beta \pd{\Psi}{x} + J\left[\Psi,\Delta\Psi\right]=0,
\end{equation}
where $\gamma$ is the inverse of the Rossby deformation length, $\rho_R$, and $\beta$ is the rate of variation of the
Coriolis parameter with latitude. The geostrophic velocity is obtained from the streamfunction
by taking the curl of $\phi\,\hat{z}$:
\begin{equation}
\v = (v_x,v_y) = \left(-\pd{\Psi}{y},\pd{\Psi}{x}\right).
\end{equation}
The Charney equation only differs from the 
two-dimensional Euler equations by the addition of the two linear terms. The physics is
somewhat different however. In particular, in addition to playing the role of the
streamfunction for the geostrophic velocity, $\Psi$ has a direct physical meaning. Because
the atmosphere is assumed to be shallow, it is proportional
to the hydrostatic pressure and, equivalently, to the deviation of the depth of the
atmosphere from its equilibrium depth. Some simple
manipulations allow it to be written as a Lagrangian conservation law:
\begin{equation}
\label{eq-PVConservation}
\frac{D}{D\,t}\left[\Delta\Psi -\gamma^2\,\Psi -\beta\,y \right] = \frac{D\,Q}{D\,t} = 0.
\end{equation}
The quantity, $Q$, which is conserved along fluid trajectories is called the potential
vorticity. 
The $\beta$ and $\gamma^2$ terms add two important ingredients to the basic dynamics described
by the Euler equations. The $\beta$ in the potential vorticity,  means that the intrinsic 
vorticity of a fluid parcel changes when it moves in the $y$ (latitudinal) direction. This
can be shown to produce a restoring force on fluid displacements in the latitudinal
direction which introduces waves into the model. These waves, known as Rossby waves are
an important feature of large scale atmospheric motions.  The $\gamma^2$ term introduces a 
potential energy penalty for the 
generation of large values of $\Psi$ or, equivalently, for large deviations of the atmospheric 
thickness from its equilibrium depth. This term sets the observed characteristic scale for
large scale motions in the atmosphere.

\subsection{Zabusky-McWilliams model}

The idea of Zabusky and McWilliams \cite{ZMcW1982} was to try to find the analogue for the
Charney equation of the point vortex representation of the Euler equations. The issue is
not straightforward since the $\beta$ term is intrinsically continuous and does not
lend itself easily to a discrete representation. The basic idea of Zabusky and McWilliams
was to start from the vortex circulation:
\begin{equation}
\label{eq-circulation}
\Omega(\x,t) = \Delta\Psi -\gamma^2\,\Psi,
\end{equation}
and discretise it on a set of $N$ time-dependent vortices, 
$\left\{\x_i(t) = (x_i(t),y_i(t)), i=1,\ldots, N\right\}$, as one would do for the
Euler case:
\begin{equation}
\label{eq-discretisation}
\Delta\Psi(\x,t) - \gamma^2\,\Psi(\x,t) = \sum_{i=1}^N \kappa_i(t)\,\delta(\x - \x_i(t).
\end{equation}
As discussed in section \ref{sec-intro}, the points $\x_i$ move with the fluid motion:
\begin{eqnarray}
\label{eq-ZM}
\dd{x_i}{t} &=& -\left.\pd{\Psi(\x,t)}{y}\right|_{\x=\x_i(t)}\\
\nonumber \dd{y_i}{t} &=& \left.\pd{\Psi(\x,t)}{x}\right|_{\x=\x_i(t)}.
\end{eqnarray}
The important difference, however, is that the vorticity is not conserved along fluid
trajectories on the $\beta$-plane. Rather the potential vorticity is conserved as in
Eq.~(\ref{eq-PVConservation}):
\begin{equation}
\label{eq-kappa_i}
\kappa_i (t) + \beta y_i(t) = \mathrm{const} = q^{(0)}_i
\end{equation}
where $q^{(0)}_i$ is the potential vorticity of vortex $i$ at $t=0$, which reveals the
dependence of the vortex strength on position.
The streamfunction is obtained from the potential vorticity by inverting the modified Helmholtz operator in
Eq.~(\ref{eq-circulation}) and using Eq.~(\ref{eq-discretisation}) and Eq.~(\ref{eq-kappa_i}):
\begin{eqnarray}
\nonumber \Psi(\x,t) &=& -\frac{1}{2\pi} \int d\y\, \Omega(\y,t)\,K_0(\gamma\, \|\x-\y\|)\\
\label{eq-streamfunction} &=& -\frac{1}{2\pi} \sum_{i=1}^N (q^{(0)}_i - \beta\, y_i(t))\, \,K_0(\gamma\, \|\x-\x_i(t)\|),
\end{eqnarray}
where $K_0(z)$ is the Bessel function of the second kind of order zero.  Eq.~(\ref{eq-streamfunction}) 
and Eq.~(\ref{eq-ZM}) now yield closed equations for the time evolution of the
vortex centres. The conservation of potential vorticity enormously changes the dynamics compared to
the case of Euler point vortices \cite{BTK2007,VFVM2003}, leading to quite complicated trajectories
even in the case of equal strength vortices.

\subsection{Zabusky-McWilliams model with a background flow}

We now discuss one way of adding a background flow, $\Psi_0(\x)$, to the Zabusky-McWilliams 
model.  We represent the contribution from a discrete set of point vortices by $\psi(\x,t)$, 
so that:
\begin{equation}
\Psi(\x,t) = \Psi_0(\x) + \psi(\x,t).
\end{equation}
Likewise the potential vorticity can be decomposed into an ambient part, $Q_0(\x)$, and a discrete part, 
$q(\x,t)$:
\begin{equation}
Q(\x,t) = Q_0(\x) + q(\x,t),
\end{equation}
where
\begin{eqnarray}
\label{eq-bgQ}Q_0(\x) &=& \Delta\,\Psi_0(\x) - \gamma^2\,\Psi_0(\x) + \beta\,y\\
\label{eq-discreteq}q(\x,t) &=& \Delta\,\psi(\x,t) - \gamma^2\,\psi(\x,t) =\sum_{i=1}^N \kappa_i(t)\,\delta(\x - \x_i(t).
\end{eqnarray}
Noting that for atmospheric flows, the total potential vorticity, $Q(\x,t)$, should be 
conserved following each vortex:
\begin{equation}
Q(\x_i(t),t) = Q(\x_i(0),0).
\end{equation}
Integrating this equation over the interior of an infinitessimal contour enclosing $\x_i$ we obtain
\begin{equation}
A\,Q_0(\x_i(t)) + \kappa_i(t) = A\,Q_0(\x_i(0))+\kappa_i(0)
\end{equation}
where $A$ is the area enclosed by the infinitessimal contour. From this, we obtain an 
approximate evolution  equation for the vortex intensities in the presence of
the background flow:
\begin{equation}
\label{eq-kappai}
\kappa_i(t) = \kappa_i(0) + A\,\left[Q_0(\x_i(0)) - Q_0(\x_i(t))\right].
\end{equation}
In Sec. \ref{sec-intro} we have noted the complication caused by the coupling between the
vortex dynamics and the evolution of the vorticity in the steering flow. Here we will
circumvent this difficulty by assuming that the circulations associated with the 
vortices are small compared to the circulation in the initial steering flow as in Eq.~(\ref{eq-ansatz}). Hence, to lowest order, the motion of the vortex centres are again obtained from Eq.~(\ref{eq-ZM}):
\begin{eqnarray}
\label{eq-mZM1}
\dd{x_i}{t} &=& -\left.\pd{\Psi_0(\x)}{y}\right|_{\x=\x_i(t)} - \left.\pd{\psi(\x,t)}{y}\right|_{\x=\x_i(t)}\\
\nonumber \dd{y_i}{t} &=& \left.\pd{\Psi_0(\x)}{x}\right|_{\x=\x_i(t)} + \left.\pd{\psi(\x,t)}{x}\right|_{\x=\x_i(t)} .
\end{eqnarray}
It remains to express the discrete part of the streamfunction, $\psi(\x,t)$ in terms of the positions
of the vortex centres. This is done, as before, from Eq.~(\ref{eq-discreteq}). We obtain
\begin{equation}
\label{eq-discretepsi}
\psi(\x,t) = -\frac{1}{2\pi} \sum_{i=1}^N \kappa_i(t) \, \,K_0(\gamma\, \|\x-\x_i(t)\|).
\end{equation}
with $\kappa_i(t)$ given by Eq.~(\ref{eq-kappai}).
The equations of motion resulting from Eqs.~(\ref{eq-mZM1}) and (\ref{eq-discretepsi}) written out
explicitly are:
\begin{eqnarray}
\label{eq-mZM2x}
\dd{x_i}{t} &=& -\left.\pd{\Psi_0(\x)}{y}\right|_{\x=\x_i(t)} - \sum_{j=1}^N \frac{\kappa_j(t)\,K_1(\gamma\left|\x_i(t)-\x_j(t)\right|)\, (y_i(t)-y_j(t))}{2\,\pi\,\left|\x_i(t)-\x_j(t)\right|}\\
\label{eq-mZM2y}
\dd{y_i}{t} &=& \left.\pd{\Psi_0(\x)}{x}\right|_{\x=\x_i(t)} + \sum_{j=1}^N \frac{\kappa_j(t)\,K_1(\gamma\left|\x_i(t)-\x_j(t)\right|)\, (x_i(t)-x_j(t))}{2\,\pi\,\left|\x_i(t)-\x_j(t)\right|},
\end{eqnarray}
where the $\kappa_i(t)$ are expressed in terms of the $\x_i(t)$ through Eq.~(\ref{eq-kappai}).
These equations differ from the original Zabusky-McWilliam model in two respects. Firstly the equations of motion
contain a velocity coming from the background flow so that a single point vortex will move following the
streamlines of the background flow even though that flow is vortical. 
Secondly, the modulation of the vortex intensities is now depends on the
background flow as specified by Eq.~(\ref{eq-kappai}). This ensures that the model remains consistent 
with the principle of conservation of potential vorticity. Clearly these equations reduce to the original
model if the background flow is absent.

We have not yet discussed possible forms for the background flow, $\Psi_0(\x)$. We require it to be a 
stationary solution of Eq.~(\ref{eq-CHM}) or, equivalently, Eq.~(\ref{eq-PVConservation}). There are
a large number of stationary solutions of the Charney equation. When $\Psi$ is independent of
time, Eq.~(\ref{eq-PVConservation}) gives
\begin{equation}
J\left[\Psi, \Delta\Psi -\gamma^2\,\Psi -\beta\,y\right] = 0.
\end{equation}
From this, is is clear that
\begin{equation}
\Delta\Psi -\gamma^2\,\Psi -\beta\,y = F(\Psi)
\end{equation}
yields a solution for any function $F(\Psi)$. In this paper, we are mostly interested in the interaction
between point vortices in the presence of a background flow rather than in the details of the background
flow itself. For this reason, we consider here two specific forms of the background flow:
\begin{enumerate}
\item
{\bf Uniform zonal current\\} 

This corresponds to a background flow consisting of a uniform westerly flow. Streamlines are straight lines.
The corresponding streamfunction is
\begin{equation}
\Psi_0(x,y) = - U\, y.
\end{equation}
\item
{\bf Inertial boundary current\\}

An inertial boundary current \cite{PED1987} occurs when a uniform westerly flow from $x=\infty$ encounters a 
straight north-south boundary at $x=0$. It is the analogue for Eq.~(\ref{eq-CHM}) of the well-known
potential solution of the two dimensional Euler equation describing a uniform flow impinging
upon a flat plate. It has a stagnation point at the origin. The streamlines are exponential curves and
the streamfunction is
\begin{equation}
\Psi_0(x,y) = U\,y\,\left[ 1 - \exp\left(-\sqrt{\frac{\beta}{U}}\,x\right)\right] .
\end{equation}
\end{enumerate}

\section{Numerical results}

\subsection{Nondimensional equations}

Let us measure lenghts in units of the Rossby deformation length, $1/\gamma$ and velocities in terms
of the characteristic velocity, $U$ of the background flow. The natural unit of time is then $(\gamma\,U)^{-1}$.
Introducing dimensionless variables, $\x^\prime$, $t^\prime$ and $\Psi^\prime$ defined by
\begin{displaymath}
\x = \frac{1}{\gamma}\,\x^\prime, \ \ \ 
t = \frac{1}{\gamma\,U}\,t^\prime, \ \ \ \mbox{and}
\Psi = \frac{U}{\gamma}\,\Psi^\prime,
\end{displaymath}
the nondimensional version of Eq.~(\ref{eq-CHM}) is (we immediately drop the primes):
\begin{equation}
\label{eq-CHM2}
\pd{}{t}\left(\Delta \Psi - \Psi\right) +\bar{\beta} \pd{\Psi}{x} + J\left[\Psi,\Delta\Psi\right]=0,
\end{equation}
where
\begin{equation}
\bar{\beta} = \frac{\beta}{\gamma^2\,U}
\end{equation}
is the dimensionless $\beta$-parameter. Our model background flows are

\begin{eqnarray}
\label{eq-zonalcurrent}
\Psi_0(x,y) &=& -  y\hspace{3.65cm} \mbox{(Uniform zonal current)}\\
\label{eq-boundaryCurrent}
\Psi_0(x,y) &=& y\,\left[ 1 - \exp\left(-\sqrt{\bar{\beta}}\,x\right)\right] .
\hspace{0.5cm}\mbox{(Boundary current)}
\end{eqnarray}

The dimensionless versions of Eqs.~(\ref{eq-mZM2x}) and (\ref{eq-mZM2y}) are
\begin{eqnarray}
\label{eq-mZM2x2}
\dd{x_i}{t} &=& -\left.\pd{\Psi_0(\x)}{y}\right|_{\x=\x_i(t)} - \sum_{j=1}^N \frac{\bar{\kappa}_j(t)\,K_1(\left|\x_i(t)-\x_j(t)\right|)\, (y_i(t)-y_j(t))}{2\,\pi\,\left|\x_i(t)-\x_j(t)\right|}\\
\label{eq-mZM2y2}
\dd{y_i}{t} &=& \left.\pd{\Psi_0(\x)}{x}\right|_{\x=\x_i(t)} + \sum_{j=1}^N \frac{\bar{\kappa}_j(t)\,K_1(\left|\x_i(t)-\x_j(t)\right|)\, (x_i(t)-x_j(t))}{2\,\pi\,\left|\x_i(t)-\x_j(t)\right|} .
\end{eqnarray}
Here the dimensionless circulations of the point vortices are
\begin{equation}
\bar{\kappa}_i(t) = \bar{\kappa}_i(0)+ \bar{A}\,\left[Q_0(\x_i(0)) - Q_0(\x_i(t))\right],
\end{equation}
with $\bar{A}$ a dimensionless area (which can be absorbed into the background flow) and $\bar{\kappa}_i(0)$
is the dimensionless initial strength of vortex $i$:
\begin{equation}
\bar{\kappa}_i(0) = \frac{\kappa_i(0)\,\gamma}{U}. 
\end{equation}
$Q_0$ is obtained from Eq.~(\ref{eq-zonalcurrent}) or Eq.~(\ref{eq-boundaryCurrent}):
\begin{equation}
Q_0(\x) = \Delta\,\Psi_0 - \Psi_0 +\bar{\beta}\,y.
\end{equation}
The upshot of all of this is that the only control parameters in the problem are the initial intensities,
$\bar{\kappa}_i(0)$, of the point vortices relative to the strength of the background flow. In what follows
we shall take all vortices to have equal initial strength. 

Some geophysically plausible values for the various parameters in the original equations are presented in
table \ref{tab-parameters}. The corresponding dimensionless values used in the numerics are summarised
in table \ref{tab-dimensionlessParameters}.

\begin{table}
\begin{tabular}{| l | c | c |}
  \hline
  Physical quantity & Value & Notes\\
  \hline
  Rossby deformation length, $\gamma^{-1}$ & $10^6 $m$$ & 1000 $km$ \\
  Beta parameter, $\beta$ & $1.6\times 10^{-11}$ $m^{-1} s^{-1}$  & \\
  Hurricane force wind velocity & 33 $m s^{-1}$ & 118 km/hr \\
  Typical radial extent of hurricane force winds & $1.6\times 10^5\ m$ & 160 km \\
  Typical hurricane eye radius & $2.4 \times 10^4\ m$ & 24 km \\
  Typical hurricane core area, $A$ & $1.81 \times 10^9\ m^2$ &  \\
  Typical circulation of a hurricane & $3.34\times 10^7\ m^2s^{-1}$ & \\
  Horizontal scale of steering flow & $5\times 10^6\ m$ & 5000 km \\
  Typical velocity of steering flow & $8 m s^{-1}$ & 17 mph \\
  Traversal time & $6.26\times 10^5\ s$& $\approx$ 7 days \\
  \hline
\end{tabular}

\vspace{0.25cm}
\caption{\label{tab-parameters} Physical values of parameters relevant to hurricane dynamics taken from \cite{NOAA1999}. Hurricane circulation has been estimated as $\Gamma = 2\,\pi\,R\,v$ based on hurricane
force winds of velocity $v=33 m s^{-1}$ extending to a distance $R=1.6\times 10^5\ m$. The vortex core
area, $A$, is estimated based on the eye radius. }
\end{table}

\begin{table}
\begin{tabular}{| l | c |}
  \hline
  Dimensionless parameter & Value \\
  \hline
  Dimensionless beta parameter, $\bar{\beta}$ & $2.0$ \\
  Initial vortex circulation $\bar{\kappa}_i(0)$ & $4.2$ \\
  Dimensionless vortex core area, $\bar{A}$ & $2.0\times 10^{-3}$ \\
  Dimensionless traversal time & 5.0 \\
  \hline
\end{tabular}
\vspace{0.25cm}
\caption{\label{tab-dimensionlessParameters} Values of dimensionless parameters used in the numerical
simulations obtained from the physical values presented in table \ref{tab-parameters}. Note 
that the vortex circulation, although large compared to $U/\gamma$ is less than that of a 
typical Atlantic anticyclone circulation ($\sim 10^8 m^2 s^{-1}$), in accordance with the
comments made before Eq.~(\ref{eq-mZM1}).} 
\end{table}

\subsection{Deflection of one vortex by another with background current}

\begin{figure}[htp]
\begin{center}
  \includegraphics[width=9cm]{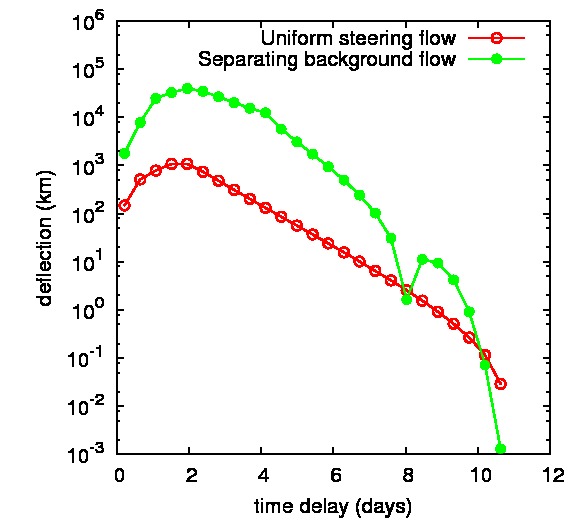}\\
  \caption{Deflection after 11 days of a vortex moving in a background current caused by a second vortex 
originating at the same location a specified time later. Open circles correspond to a uniform backround current, 
solid circles to a separating background flow.}
  \label{fig-deflection}
  \end{center}
\end{figure}

\begin{figure}[htp]
\begin{center}
\begin{tabular}{cc}
  (A) Unperturbed track & (B) Delay of 0.5 days\\
  \includegraphics[width=5cm]{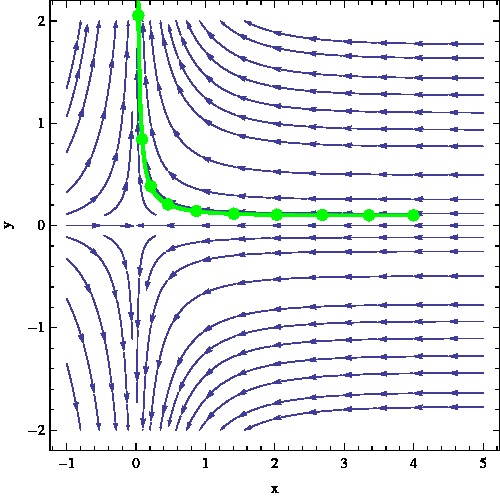}& \includegraphics[width=5cm]{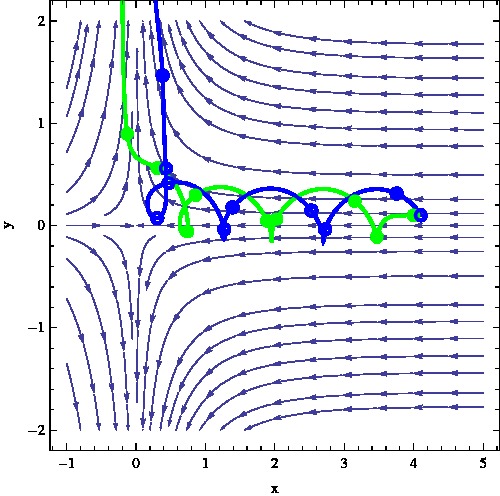}\\
  (C) Delay of 1 day & (D) Delay of 2 days\\
  \includegraphics[width=5cm]{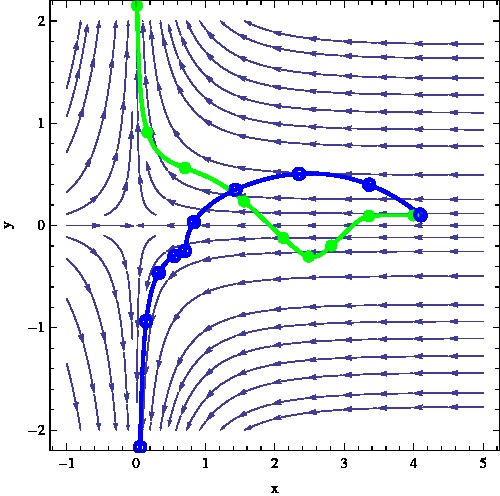}& \includegraphics[width=5cm]{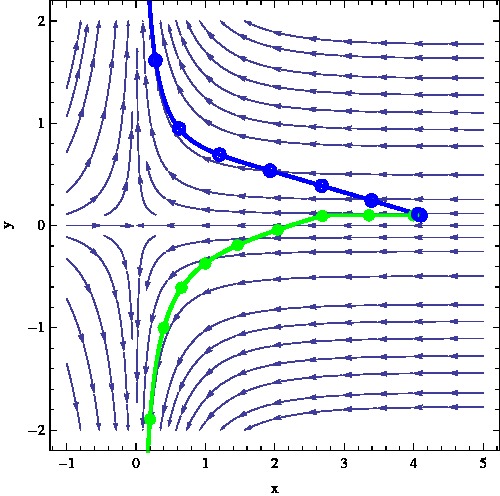}\\
  (E) Delay of 3 days & (F) Delay of 4 days\\
  \includegraphics[width=5cm]{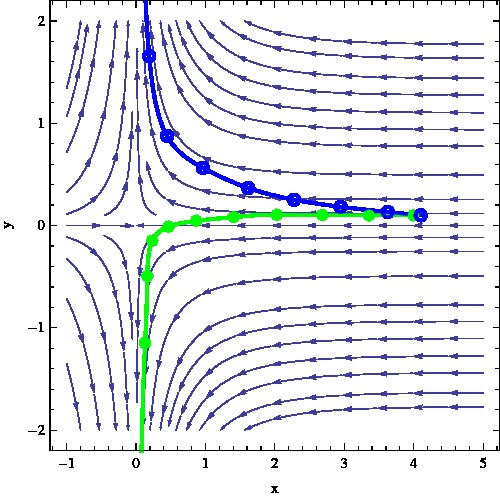}& \includegraphics[width=5cm]{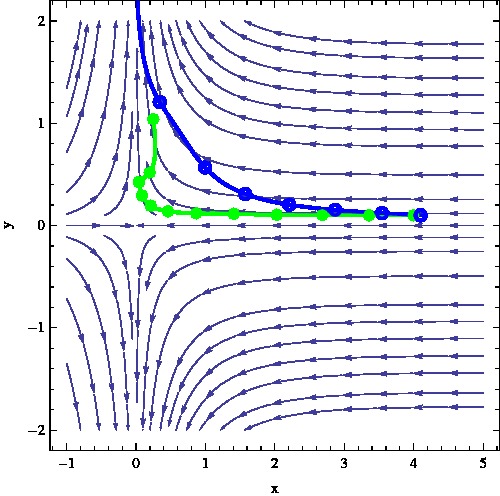}
\end{tabular}
  \caption{Tracks of vortices in a separating background flow for different delay times. Panel (A)
shows the unperturbed track in the absence of a second vortex. Very large deflections, of the
order of the system size, are observed for delays of up to 4 days.}
  \label{fig-tracks}
  \end{center}
\end{figure}

Fig.~\ref{fig-deflection} shows the results of a numerical experiment designed to quantify the effects 
of vortices on each other. The parameters used are as in Table \ref{tab-parameters} and the axes have
been re-expressed in physical units for ease of interpretation. We used two different background flows, 
a uniform current flowing to the left (west) (open circles) and separating background flow 
corresponding to an inertial boundary current (solid circles). 
A single point vortex  simply follows the streamlines of the background flow as it
would have done in the classical theory of potential flows. In order to quantify the influence
of other vortices on this trajectory we started a vortex off at a 
chosen point in space and then inserted a second vortex at the same point after a given time delay of $\tau$.
We then compared the position of the first vortex at a later time, $T=11$ days, with the position it would have
been at had the second vortex not been inserted. The follow-my-leader effect can then be quantified by
plotting this deflection of the first vortex from its unperturbed path as a function of the time delay between the vortices. Clearly, for
large delays, the deflection should be smaller as the time delay increases. 

For the case of a uniform 
background flow, it decreases exponentially as the delay increases. We see that for a delay of 7 days,
typical for Atlantic hurricanes, the deflection is about 10 km. Since this is less than the core radius of
our vortices, below which our model cannot be considered meaningful, this means that there is effectively
no interaction between the vortices for such separations. Of course, much stronger deflections of the order
of hundreds of km can be observed for vortices separated by 2 days or less which is reflecting the well-known
fact that vortices interact very strongly when they get close to each other.

For the separating background flow, the typical deflection is orders of magnitude larger even though all
model parameters remain the same. The reason is clear from Fig. \ref{fig-tracks} which illustrate the 
tracks followed by the two vortices for different values of the delay. One sees that the second vortex can
easily deflect the first one onto a track which subsequently diverges from the initial track due
to the presence of the stagnation point in the flow.

\section{Conclusion}

We have used a quasi-geostrophic model together with a physically plausible law of vortex
dynamics to model hurricane tracks in the presence of an ocean-scale steering flow. Our
model has enabled us to quantify the sensitivity of the ``follow-my-leader'' phenomenon to
the presence of stagnation points in the steering flow. It would be of interest in the
future to compare the predictions of our simulation with the probabilistic models that are
used in risk estimation. 

\bibliographystyle{plain}

\begin{thebibliography}{1}

\bibitem{BTK2007}
I.~J. Benczik, T.~T\'{e}l, and Z.~K\"{o}ll\"{o}.
\newblock Modulated point-vortex couples on a beta-plane: dynamics and chaotic
  advection.
\newblock {\em J. Fluid Mech.}, 582:1--22, 2007.

\bibitem{BEYL1998}
W.~{Bin}, R.~L. {Elsberry}, W.~{Yuqing}, and W.~{Liguang}.
\newblock Dynamics in tropical cyclone motion: a review.
\newblock {\em Chinese J. Atm. Sci.}, 22(4):416--434, 1998.

\bibitem{CHA1949}
J.~G. Charney.
\newblock On a physical basis for numerical prediction of large-scale motions
  in the atmosphere.
\newblock {\em J. Meteor}, 6:371--85, 1949.

\bibitem{LH1993}
M.~{Lander} and G.~J. {Holland}.
\newblock {On the interaction of tropical--cyclone--scale vortices. I:
  Observations}.
\newblock {\em Quart. J. Roy. Met. Soc.}, 119:1347--1361, 1993.

\bibitem{ESGI73}
L.~MacManus et~al.
\newblock {Modelling Hurricane Track Memory}.
\newblock Report of 73rd European Study Group with Industry,
  http://www.maths-in-industry.org/miis/view/studygroups/esgi73/, 2010.

\bibitem{NOAA1999}
NOAA.
\newblock {Hurricane Basics}.
\newblock http://hurricanes.noaa.gov/pdf/hurricanebook.pdf, 1999.

\bibitem{PED1987}
J.~Pedlosky.
\newblock {\em Geophysical fluid dynamics, 2nd Ed.}
\newblock Springer, New York, 1987.

\bibitem{VFVM2003}
O.~U. {Velasco Fuentes} and F.~A. {Vel\'{a}zquez Mu\~{n}oz}.
\newblock Interaction of two equal vortices on a $\beta$-plane.
\newblock {\em Phys. Fluids}, 15(4):1021--1032, 2003.

\bibitem{ZMcW1982}
N.~J. {Zabusky} and J.~C. {McWilliams}.
\newblock A modulated point-vortex model for geostrophic $\beta$-plane
  dynamics.
\newblock {\em Phys. Fluids}, 25(12):2175--2182, 1982.

\end{thebibliography}

\medskip
Received xxxx 20xx; revised xxxx 20xx.
\medskip

\end{document}